\documentclass[aps,twocolumn,showpacs]{revtex4}

\usepackage{bm}
\usepackage{graphicx}
\usepackage{amsmath}
\usepackage{eufrak}

\usepackage{color}

\usepackage{ulem}

\usepackage{graphics}

\usepackage{dcolumn}

\begin{document}

\title{Time resolution and dynamic range of field effect transistor based terahertz detectors}

%

\author{Przemyslaw Zagrajek$^1$,  Sergey N. Danilov$^2$,
    Jacek Marczewski$^3$,  Michal Zaborowski$^3$,  Cezary Kolacinski$^3$,  Dariusz Obrebski$^3$,
    Pawel Kopyt$^4$, Bartlomiej Salski$^4$, Dmytro But$^5$, Wojciech Knap$^5$,  and Sergey D. Ganichev$^2$}
\affiliation
{$^1$ Institute of Optoelectronics, Military University of
Technology, Warsaw, Poland\\
%
$^2$ Regensburg Terahertz Center (TerZ), University of Regensburg,
Germany\\
$^3$ Institute of Electron Technology, Warsaw, Poland\\
$^4$ Inst. of Radioelectronics and Multimedia Technology, Warsaw
University of Technology, Poland\\
$^5$ International Research Centre CENTERA, Institute of High
Pressure Physics, Polish Academy of Sciences, Warsaw, Poland }




\begin{abstract}
We studied time resolution and response power dependence of three
terahertz detectors based on significantly different types of
field effect transistors. We analyzed the photoresponse of
custom-made Si junctionless FETs, Si MOSFETs and GaAs-based high
electron mobility transistors detectors. Applying monochromatic
radiation of high power, pulsed, line-tunable molecular THz laser,
which operated at frequencies in the range from 0.6-3.3~THz, we
demonstrated that all these detectors have at least nanosecond
response time. We showed that detectors yield a linear response in
a wide range of radiation power. At high powers the response
saturates varying with radiation power $P$ as $U = R_0
P/(1+P/P_s)$, where $R_0$ is the low power responsivity, $P_s$ is
the saturation power. We demonstrated that the linear part
response decreases with radiation frequency increase as $R_0
\propto f^{-3}$, whereas the power at which signal saturates
increases as $P_s \propto f^3$. We discussed the observed
dependences in the framework of the Dyakonov-Shur mechanism and
detector-antenna impedance matching. Our study showed that FET
transistors can be used as ultrafast room temperature detectors of
THz radiation and that their dynamic range extends over many
orders of magnitude of power of incoming THz radiation. Therefore,
when embedded  with current driven read out electronics they are
very well adopted for operation with high power pulsed sources.
\end{abstract}


\maketitle

\section{Introduction}

Photovoltaics based on plasma oscillations
is considered as the main candidate for robust and highly
sensitive room temperature THz photodetectors, see
e.g.~\cite{x3,Vicarelli,x1,x2,x4,x5}. Plasmons are high-frequency
oscillations of the electron gas density and occur in many metals
and semiconductors. A boost to the plasma wave semiconductor
electronics was given at early 90s, when M. Dyakonov and M. Shur
predicted a $dc$-current-induced plasma wave instability in
nanometer sized channels of field-effect transistors (FET) and
demonstrated that FET can be used for efficient detection of
microwave/terahertz radiation~\cite{Dyakonov,Dyakonov2}. The
instability results in the generation of plasma oscillations. Due
to the high velocity of these plasma waves ($>10^8$ cm/s) the
frequency of the oscillations can be easily tuned (e.g., applying
a gate voltage) into the THz frequency
range~\cite{Dyakonov,Dyakonov2}. Furthermore, in high mobility
structures plasmonic resonances result in a drastic increase of
the \textit{dc} signal response to THz radiation. Robust room
temperature FET plasmonic detectors being characterized by large
detectivity (D$^*$) and, correspondingly, low noise equivalent
power (NEP) has become an important technology for THz
applications. Main nowadays demonstrated THz technology
applications are nondestructive quality control or security
screening. In case of highly absorbing media (opaque or very thick
materials/packages) imaging with CW sources  suffers from the
quickly degrading signal to noise ratio. Indeed, while using
continuous wave  THz sources  even use of the most sensitive
detectors does not allow to get sufficiently high image quality.
This is because existing CW sources operate in relatively low
output power: milli- or microwatt range.  This can be re-mediated
by use of  of high-power terahertz sources such as molecular
lasers, free-electron lasers, $p$-Ge lasers
difference-frequency-based terahertz systems, impurity laser in
stressed bulk and low dimensional semiconductors, etc., see
e.g.~\cite{book,Lee,Zhang,Bruendermann,Elsaesser},
besides D$^*$ and NEP. These sources can reach powers up to a few
watts however they  always operate in the pulsed mode. Therefore,
while studying FETs for imaging applications with these sources
one has to consider not only FET responsivities but also
additional  figures  of merit like time resolution,  and dynamic
range. The later has been rarely addressed in the previous studies
but is really important in the imaging. This is because  during
scanning of highly attenuating media/objects, with high power
sources, the  power of  THz beam reaching detector increases by
many orders of magnitude while reaching the boarder of the object.
This often leads even to breakdown   of many of the standard
detectors like for example Schottky diodes. On the other hand many
of the robust standard detectors like for example Si-bolometers
are not fast enough to be used with high-power pulsed sources.
Former characteristic is also important for wireless
communication, for which FET are considered as a promising
candidate because this solution guarantees low fabrication costs
and easy on-chip integration with potential readout electronic.
The purpose of this work is to study FET as detectors for high
power short pulsed THz sources.

Here we present a systematic study of these characteristics of
three different types of the state-of-the-art plasma based
detectors. We analyse photoresponse of a custom-made detector Si
junctionless FET (JLFET), Si metal oxide semiconductor FET
(Si-MOS) and InGaAs/GaAs-based high electron mobility transistors
(HEMT) transistors, which represent structures that are very often
employed in experiments dedicated to detection of sub-THz and THz
radiation using FETs. All structures have been monolithically
integrated with a log-periodic broad-band planar antenna. Time
resolved photoresponse of room temperature detectors have been
studied applying nanosecond pulses of monochromatic linearly
polarized terahertz laser radiation with frequencies $f$ ranging
from 0.6 to 3.3 THz. Our measurements explore time resolution,
detectors nonlinearity, and polarization dependence for different
gate voltages applied to FET transistors operating with two types
of amplifiers. We show that indeed FET detectors, while embedded
with specific read-out circuits (current driven) are well adapted
(low NEP, high dynamic range) for high power pulsed laser based
applications.

\subsection{Devices and characteristics}

\begin{figure}[h]
    \center
    \includegraphics[width=0.6\linewidth]{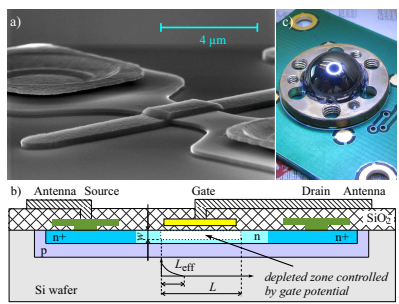}
    \caption{JLFET transistor. (a) -- Typical scanning
    electron microscopy micrograph of the device;  (b) -- Cross-section;
    (c) -- Photo of the device with integrated silicon lens
    (in the middle of the front panel) for focusing the incident
    radiation.
    } \label{fig01}
\end{figure}

Figure~\ref{fig01} shows scanning  electron microscopy image and
sketch of the first structure: Junctionless FET fabricated  in
Institute of Electron Technology, Poland. JLFET is a device with a
layout that is similar to classical MOSFET one, however, in
contrast to classical MOSFETs, the source, drain and channel areas
of JLFET are made of the same type of semiconductor ($n^+$- and
$n$-type), see Fig.~\ref{fig01}(b). The JLFET detector has been
fabricated on high resistivity silicon wafer. Regions of $p$-type
well and $n$-type channel were subsequently diffused. Further
processing steps included gate thermal oxidation (thickness
25~nm), poly-Si gate preparation (thickness 400~nm), and Al:Si
metal contact fabrication. The log-periodic antenna described
below was also made of this metal layer. The source-drain current
in JLFET can be varied by the thickness of the conducting channel
below the depleted zone controlled by the gate voltage $V_g$. The
$n^+$-$n$-$n^+$ conduction path has $n$-area of carrier density
$2.5 \times 10^{16}$~cm$^{-3}$, which has been choosen to control
the conduction channel by gate voltages within a reasonable range.
Note that our junctionless FET remains conductive even without the
gate voltage and closes at -0.4~V threshold voltage $U_{\rm th}$.
To minimize any parasitic gate-source and gate-drain capacitances
the polysilicon gate was deposited at about $1~\mu$m distance from
$n^+$ source and drain regions, see Fig.~\ref{fig01}(b).  The
detector gate length and width were 5 and $12~\mu$m, respectively.
The carrier mobility determined from accumulation characteristics
was $\mu_n$ = 154 cm$^2$/(V$\cdot$s) near the transistor's
threshold voltage $U_{\rm th}=-0.4$~V. The effective length
$L_{\rm eff}$,  where rectification takes place,  was estimated
according to Ref.~\cite{Kachorovskii2013}. For gate voltage $ U_g
= 0$ and frequency $f = 1$~THz we obtained  $L_{\rm eff} =31$~nm.
More details on the structure preparation as well as electrical
characteristics of the similar structures can be found in
Ref.~\cite{Marczewski2015,ProcSPIE2016}.

\begin{figure}[h]
    \center
    \includegraphics[width=0.6\linewidth]{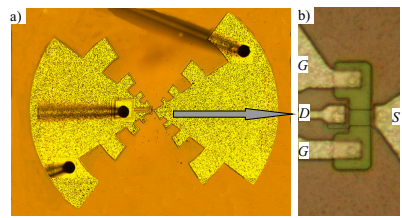}
    \caption{Typical design of the sample antenna (a) with a zoomed central part (b).  }
    \label{fig02}
\end{figure}

Second structure  studied here was silicon MOS transistor, also
made in Institute of Electron Technology, Poland. The structure
had a classical $n$-type MOSFET design and was fabricated applying
CMOS technology ($3~\mu$m design rules, poly-Si gate) modified to
use a high-resistive substrate, which in this way became part of a
glued  hyper-spherical silicon lens described below. The gate
length and width were 3 and  6~$\mu$m, respectively.  The carrier
mobility was {$\mu_n$ = 530 cm$^2$/(V$\cdot$s)} near the threshold
$U_{\rm th}=0.91$~V ($L_{\rm eff} =96$~nm).

The third structure was HEMT transistor based on epitaxially grown
InGaAs/GaAs structure, which have been designed and fabricated
within the PH10 GaAs pHEMT process offered commercially by United
Monolithic {Semiconductors S.A.S} (UMS).    The channel
conductivity could be controlled by the gate voltage from -1.0 to
0.7~V. The gate length was $L = 0.1$~$\mu$m, the gate width was
from 20~$\mu$m. More details on the HEMT characteristics can be
found in Ref.~\cite{Teyssandier}.

Each of these transistors has been monolithically integrated with
a  broadband log-periodic planar antenna located on the top-most
metallization layer, see Fig.~\ref{fig02}. The antenna directly
feeds the source and the gate contacts of the transistor. The
metallization layer is also used to provide the DC-bias voltages
to the structure. The $U_{sd}$ voltage is accessible at contact
pads located near the outer edge of the lower arm of the antenna.
The source electrode was grounded and the output signal, $U$, was
detected from a drain electrode. The channel resistance (and
detectors responsivity) was controlled by applying a gate voltage
$U_g$. The complete device has been designed as a broadband
detector, which requires HR high-resistivity hyper-spherical
lens~\cite{Filipovic}.  The detector elements has been mounted on
the flat side of the lens of diameter 10 mm, see  Fig.~\ref{fig01}
(c).

\subsection{Sources of radiation and methods}

Detectors responsivity and noise equivalent power have been
measured applying low-power $cw$ radiation of amplifier-multiplier
chain (AMC-10 unit from VDI, Inc.). This source provided
monochromatic incident wave of frequency selectable in the band of
0.14 – 0.5 THz. Signals were detected using a lock-in amplifier
(SR-830 from Stanford Research Systems, Inc.) synchronized with a
low-frequency chopping signal that was modulating the THz source
during measurements. The modulation frequency was equal 187~Hz.
The beam profile was nearly Gaussian. Measurements were performed
without any additional beam shaping elements. The distance between
the source and investigated devices was about 30~cm. An estimation
of the responsivity given in V/W was obtained by comparison with
calibrated detector of the same size, for the details of the
method see Ref.~\cite{Kopyt}. The method, which was  used instead
of calculation power deposited on a detector aperture, allowed us
to decrease large systematic error. As a calibrated device, a
commercial quasi-optical Schottky diode detector (QOD 1-14 from
VDI, Inc.) was used. To calculate NEP per unit bandwidth the
Johnson-Nyquist noise was assumed as dominating and took into
account.

\begin{table}[h]
    \centering
    \begin{tabular}{|c|c|c|c|c|}
        \hline
        &Si MOS                   &JLFET               & HEMT  \\
        \hline
        Responsivity (V/W)        &6                   & 0.5             & 150    \\
        NEP (W/$\sqrt{\rm{Hz}}$)  &$7.2\times10^{-9}$  &$3\times10^{-8}$ & $1.3 \times10^{-11}$ \\
        \hline
    \end{tabular}
    \caption{Responsivity and noise equivalent power measured for $f=0.18$~THz. }
    \label{tab03}
\end{table}

Time resolution and detectors nonlinearity have been measured by
applying high-power line-tunable pulsed molecular gas THz
laser~\cite{ratchet2011,Drexler2012} optically pumped by a tunable
CO$_2$ laser~\cite{PRB2002,PRBratchet}. Using CH$_3$F, D$_2$O, and
NH$_3$ as active media, monochromatic  laser radiation with
frequencies of 0.6, 0.77, 1.07, 2.02, and 3.3\,THz  were obtained.
The laser operated in single pulse regime (repetition frequency
1~Hz) with a pulse duration of about 100\,ns. Due to the
spontaneous mode-locking the pulse consisted of short spikes with
full width half maximum time duration of about 4~ns, which allowed
us to analyse time constant of detectors with a a resolution of a
several ns.  The radiation power $P$ has been measured by a
calibrated fast room temperature  photon drag
detector~\cite{Ganichev85p20} and $\mu$-photoconductivity
detector~\cite{Ganichev85p377}, both made of  $n$-type Ge
crystals. These  detectors operating in the frequency range for
0.3 to 30 THz, had subpicosecond time resolution and were used as
reference detectors for the time resolved measurements. For both
reference detectors we used a bandwidth of 300 MHz and a voltage
amplification of 46~dB.
The radiation was focused by a parabolic mirror. The laser beam
had an almost Gaussian shape with full widths at half maximum
between 1 mm (at 3.3~THz) and 3 mm (at 0.6~THz) as measured by a
pyroelectric camera~\cite{PhysicaB1999}.  The highest peak power
and power density used in our measurements at these frequencies
were 1~kW and 0.1~W/cm$^2$, respectively. Note that in the used
laser peak power varies from pulse to pulse by about 15\,\%. All
measurements have been carried out at room temperature and the
devices have been illuminated at normal incidence. To vary the
radiation intensity we used a set of teflon, black polyethylene,
polyoxymethylene, polyvinyl chloride, and/or pertinax~\cite{book}
calibrated attenuators. The photoresponse was amplified and
measured using a digital oscilloscope. Two types of amplifiers
with substantially different input resistance  have been used. The
first one was transimpeadance amplifier (TIA) with very small
input impedance ($< 1~\Omega$), whereas the second one was voltage
follower (VF) or, to be more precise, unity-gain voltage
amplifier. Amplifiers characteristics are given in
Tab.~\ref{tab02}.

\begin{table}[h]
    \centering
    \begin{tabular}{|c|c|c|c|c|}
        \hline
        & Bandwidth & input resistance & output resistance      & amplification   \\
        &  (MHz)    & DC ($\Omega$)    & at  10 MHz ($\Omega$)  &                 \\
        \hline
        VF  &    108        &  450 k$\Omega$  &       2         &  1 V/V          \\
        TIA &    470        &   $ < 1\Omega$  &      50         & 8000 V/A         \\
        \hline
    \end{tabular}
    \caption{Characteristics of the used amplifiers. Note that internal limits
    of VF and TIA amplifiers are 1.6 and  0.3~ns, respectively. }\label{tab02}
\end{table}

The voltage follower  is a circuit providing high resistance of
its input and relatively low resistance of the output. This
circuit reduces the influence of the capacitance of cables
connecting the detector under examination with the lock-in system
at the expense of the amplitude of the signal. At  the same time
VF meets the assumption of the Dyakonov theory, which requires a
detector to be loaded with a high resistance (open circuit). The
transimpedance amplifier  is an amplifier whose input current
produces an output voltage in a proportional way. Ideally its
input should represent a short circuit. Certainly such a circuit
is not commonly used for THz radiation detection with FET
detectors but allows to avoid a limitation of transmitted
bandwidth associated with the output capacitance of the detector.
Therefore, such an idea is often used for design of optical
amplifiers for telecommunication technology.

Additional measurements  on power dependence have been carried out
applying CW CH$_3$OH optically pumped
laser~\cite{Olbrich2013,Dantscher2015} operating at frequency of
2.54~THz. The laser beam with power $P= 60$~mW was modulated at a
frequency of $f_{\rm chop} = 36$~Hz. Note that TIA has $-3$~dB low
frequency cutoff at modulation frequency about about 30 kHz,
therefore, at our low $f_{\rm chop}$ its amplification coefficient
was substantially reduced.

\section{Results and discussion}

\begin{figure}[h]
    \center
    \includegraphics[width=\linewidth]{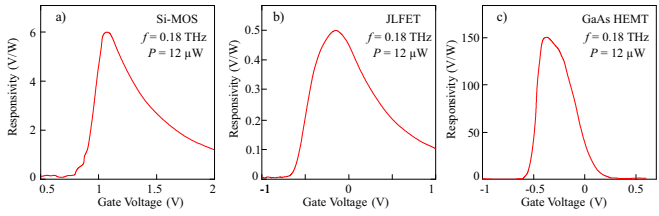}
    \caption{Responsivity of the detector element (before amplification)
    as a function of a gate voltage  measured  with the amplifier-multiplier
    chain AMC-10 unit from VDI, Inc. operating at $f=0.18$~THz and power of 12~$\mu$W.
    (a) -- for silicon MOS sample; (b) -- for silicon JLFET structure; (c) --  for InGaAs/GaAs HEMT sample.
        } \label{fig03}
\end{figure}

\begin{figure}[h]
    \center
    \includegraphics[width=\linewidth]{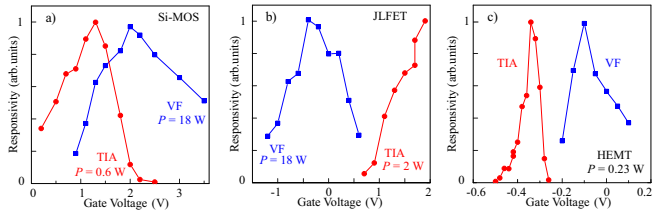}
    \caption{Gate voltage dependencies of the photoresponse to the
    radiation with $f=1.07$~THz for detectors with different
    electronic measurement circuits (TIA -- transimpedance amplfier,
    VF -- voltage follower) (a) -- for Si-MOS sample; (b) -- for
    silicon JLFET structure; (c) -- for HEMT. Points are
    joint by lines only for eye.
} \label{fig04}
\end{figure}

\begin{figure}[h]
    \center
    \includegraphics[width=\linewidth]{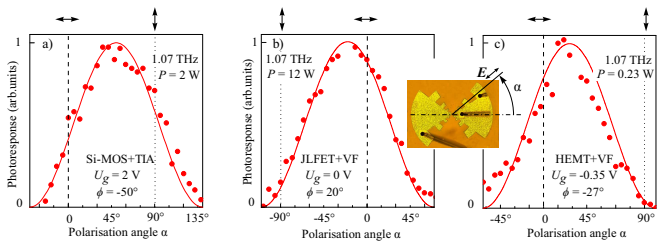}
    \caption{Photoresponse as a function of the azimuth angle $\alpha$
    defining the polarization direction of the linearly polarized radiation
    in respect to the $x$-direction. (a) Si-MOS; (b) silicon JLFET and (c) HEMT.
    Lines are fit after $U = U_m \cos^2 (\alpha  + \phi)$. Note that for clarity
    we subtracted from the data a small offset $U_{\rm off}$ being less
    than 10\% of the amplitude $U_m$. The phase shifts, $\phi$, used for fits
    are $-50^\circ$ (Si-MOS),  $20^\circ$ (JLFET) and  $-27^\circ$
    (HEMT). For zero angle $\alpha$ the radiation electric field vector is parallel
    to the $x$-axis. Arrows on top of panels illustrate the orientations
    of the radiation electric field vector for $\alpha =0$ and 90$^\circ$.
    } \label{fig05}
\end{figure}

We begin with briefly introducing  the results obtained by
applying low frequency and low  power radiation of  $cw$ radiation
of amplifier-multiplier chain zero magnetic field. While the paper
is devoted to detector's time resolution and nonlinearity  these
results show basic characteristics of our detectors obtained under
''standard'' experimental conditions. Results of experiments with
linearly polarized low power radiation are shown in
Fig.~\ref{fig03}. The figure presents a gate voltage dependence of
the transistor's photoresponse measured for radiation frequency of
$f=0.18$~THz and power 12~$\mu$W. For all transistors the signal
shows a maximum for a gate bias close to the threshold voltage,
$U_{\rm th}$. Such a non-monotonic behavior of the signal is well
known for FET
detectors~\cite{Knap2002,Sakowicz2011,Lusakowski2011}.
Responsivities $R_0$ and estimated noise equivalent powers NEP are
given in Tab.~\ref{tab03}. The values of $R_0$ are smaller and NEP
are larger than the best found in literature. In our opinion the
reason is that our method of responsivity calculation does not
include a size of detector.

Photoresponse of all three detectors  has been obtained by
applying pulsed radiation with substantially higher frequencies,
including the highest one used in our work, $f=3.3$~THz.
Figure~\ref{fig04} shows  gate voltage dependencies obtained for
$f=1.07$~THz. Dependencies were measured with two different
amplifiers: the TIA and the VF. While in both cases we obtained
nonmonotonic dependencies  similar to that measured at
$f=0.18$~THz, see Fig.~\ref{fig03}, the maximum position is
substantially different for TIA and VF amplifiers. We attribute
this fact to more than five order difference in the input
resistance, see Tab.~\ref{tab02}. These observed differences in
location of the maximum responsivity point, as a function of gate
voltage,  are related to totally different operating point of
detecting transistors. The output current of a FET is also
affected by its drain to source voltage. In case of the TIA, the
output electrode (drain) of the detecting transistor is loaded by
a very low input resistance of this read-out amplifier. In
contrast to TIA, the VF has input resistance ranging to hundreds
of kiloohms. Varying orientation of the radiation electric field
vector by means of lambda half plates we observed that the signal
follows $U = U_m \cos^2 (\alpha  +\phi)$ dependence, where  $U_m$
is the photoresponse magnitude and $\phi$ is a phase shift.  Such
polarization dependence is expected for Dyakonov-Shur FET
detectors additionally taking into account the the optimum
polarization for radiation-antenna coupling, see
e.g.~\cite{Sakowicz2008,Sakowicz2010,antenna2005}. Note that in
some measurements we also detected a small polarization
independent offset.  Comparison of the photovoltage distribution
with the design of  transistors reveal that in JLFET and HEMT
detectors the signal achieved a maximum value for the radiation
polarization vector aligned nearly parallel to the line connecting
the source and drain, being also axis of the antennas, see
Fig.~\ref{fig02}. In a Si-MOS structure we detected the maximum of
the response at an angle about 45$^\circ$ to this axis.

Now we turn to the main part of the paper devoted to time
resolution, frequency range and nonlinearity of the FET detectors.

\begin{figure}[h]
    \center
    \includegraphics[width=\linewidth]{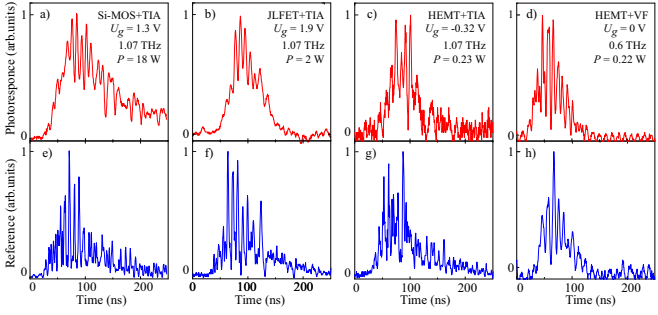}
    \caption{Temporal structure of the photoresponse. (a)-(c) Transistors
    combined with low input resistance amplifier TIA and (d) HEMT combined
    with high input resistance amplifier VF. Note that all pulse traces
    were obtained at power level corresponding to the linear regime of
    detectors operation. (e)-(h) Traces measured with reference fast photon drag detector.
    } \label{fig06}
\end{figure}

\begin{figure}[h]
    \center
    \includegraphics[width=\linewidth]{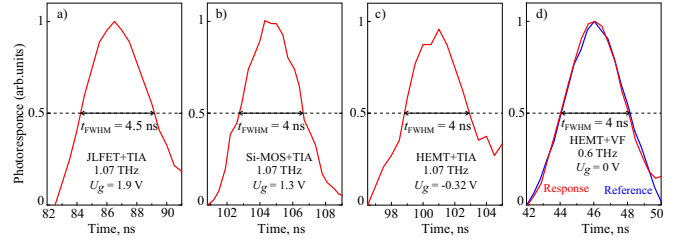}
    \caption{Temporal structure of photoresponces showing zoom of
    selected from Fig.~\ref{fig06} single short spikes caused by the
    spontaneous mode-locking. (a)-(c) Transistors combined with low
    input resistance amplifier TIA and (d) HEMT combined with high
    input resistance amplifier VF together with the signal of reference
    photon drag detector. Note that all pulse traces were obtained at
    power level corresponding to the linear regime of detectors operation.
    }
 \label{fig06bis}
\end{figure}

Figures~\ref{fig06}(a)-(c) show  pulse traces obtained with FET
detectors combined with transimpeadance amplifier characterized by
its input resistance less than $1 \Omega$ in comparison with the
response of the reference photon drag detector. In this case the
output terminal (drain) potential was pinned to 0. Under this condition, the response
of any detector is usually much faster than if it was loaded with
an amplifier with the large input resistance, when RC constant
regarding the resistivity of its channel and the output
capacitance plays a dominant role. Figure~\ref{fig06}(d) proves,
however, that the extremely small RC constant typical of modern
sub-micron HEMT transistors makes them sufficiently fast, even
when they are loaded with the high VF resistance. It is seen that
with TIA amplifiers all detectors perfectly reproduce short spikes
caused by the spontaneous mode-locking. The same is valid for HEMT
at zero gate voltage combined with VF amplifier. Zoom of single
spikes are shown in  Fig.~\ref{fig06bis}. The full width half
maximum (FWHM) time duration, $t_{\rm FWHM}$, of all signals
registered by FET detectors is about 4~ns. The same time constants
are detected by fast reference photon drag detector, see an
example in Fig.~\ref{fig06bis}(c), and $\mu$-photoconductivity
detectors (not shown). Consequently, our measurements demonstrate
that all three detectors have time resolution better than 4~ns.
This result has been obtained for all gate voltages and radiation
frequencies. Such a high time resolution of FET detectors is in
fully agreement with theoretical results of
Ref.~\cite{Kachorovskii2008}. This paper considered the dynamic
response of FET for modulation frequencies lower than radiation
frequency. It was shown that theoretical the maximal modulation
frequency allowing for adiabatic response is rather high (on the
order of 50–100 GHz for a 200~nm gate transistor operating
above-threshold and on the order of 5–10 GHz in the
below-threshold regime).

\begin{figure}[h]
    \center
    \includegraphics[width=\linewidth]{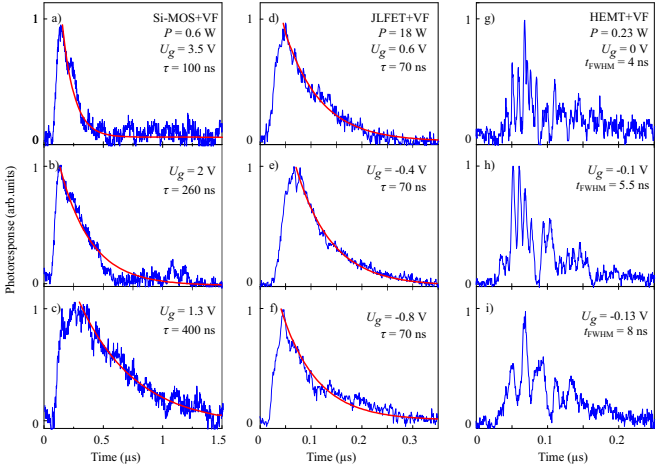}
    \caption{Temporal structure of the photoresponse measured with FET
    combined with high input resistance voltage follower VF. Pulse
    traces are obtained for different gate voltages. (a)-(c) Si-MOS;
    (d)-(f) JLFET and (g)-(i) HEMT. Note that all pulse traces were
    obtained at power level corresponding to the linear regime of detectors
    operation. Curves in panels (a)-(f) show fits after $U\propto exp(-t/\tau)$,
    where decay time $\tau$ is the fitting parameter. Times $t_{\rm FWHM}$ given
    for HEMT transistors are obtained from analysis of single short spikes
    caused by the spontaneous mode-locking using procedure described in Fig.~\ref{fig06bis}. }
    \label{fig07}
\end{figure}

\begin{figure}[h]
    \center
    \includegraphics[width=0.5 \linewidth]{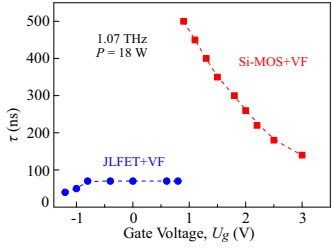}
    \caption{Decay time $\tau$ as a function of the gate voltage. The data
    are obtained from the pulse traces measured by Si-MOS and JLFET with
    the voltage follower VF. Lines are guide for eye.
    } \label{fig10}
\end{figure}

Using the  same detector elements with voltage follower we
observed a substantial increase  of the response time of Si-MOS
and JLFET  as well as a small increase of the one for HEMT.
Furthermore, it becomes dependent on the gate voltage. Typical
pulse traces obtained with VF amplifiers are shown in
Fig.~\ref{fig07}. Figures~\ref{fig07}(a)-(f) show that both
silicon FETs loaded with VF are much slower comparing to those
loaded with TIA, see Figs. ~\ref{fig06} and~\ref{fig06bis}. They
hardly follow the envelope of the signal and definitely cannot
reproduce individual spikes well seen on Fig. ~\ref{fig06} as
measured by the photon drag detector. In VF configuration the $RC$
constant of these detectors becomes the most significant factor
limiting their speed.  Figures ~\ref{fig07}(a)-(c) show that
opening channel of Si-MOSFET (going from 1.3 toward 3.5~V)  makes
the device faster since the channel opens proportionally to the
$U_g-U_{\rm th}$ difference reducing the resistivity of the
channel. A more than 20~times increase of the response time is
also detected for JLFET, see Figs.~\ref{fig07}(d)-(f). In the case
of JLFET, the gate bias between -0.8 and 0.6~V has slight
influence on the carrier amount in the channel, thus the time
constant does not change with the gate voltage. Figure~\ref{fig10}
shows the decay time as a function of the gate voltage for both
Si-FETs. Figures~\ref{fig07}(g)-(i) proves, however, that
sub-micron HEMT is a much faster device characterized by extremely
small $RC$ constant (the individual spikes seen by the drag
detector are well seen). In this case opening channel (changing
$U_g$ from -0.13~V to 0) makes the device only a bit faster.

\begin{figure}[h]
    \center
    \includegraphics[width= \linewidth]{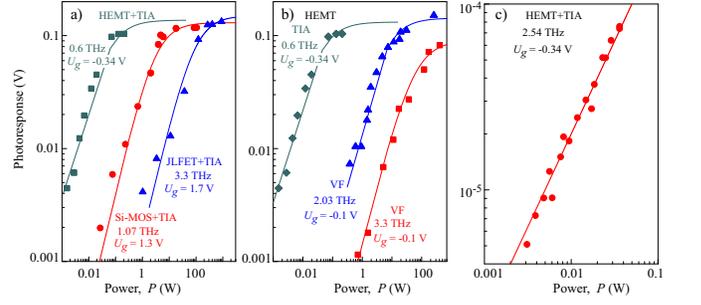}
    \caption{Power dependence of the photoresponse. (a) Si-MOS,
    JLFET and HEMT transistors combined  with TIA amplifier
    (fast response regime); (b) Data for different radiation frequencies
    obtained by HEMT combined with TIA and VF amplifiers; (c) results
    for CW-laser and HEMT transistor. Lines are fit after $U = R_0 P / (1 + P/P_s)$.
    The data are presented for gate voltages corresponding to the maximum
    of photoresponse, see Fig.~\ref{fig05}, and for fixed polarization of radiation
    with electric field vector along $x-$axis.
    }
    \label{fig08}
\end{figure}

\begin{figure}[h]
    \center
    \includegraphics[width=\linewidth]{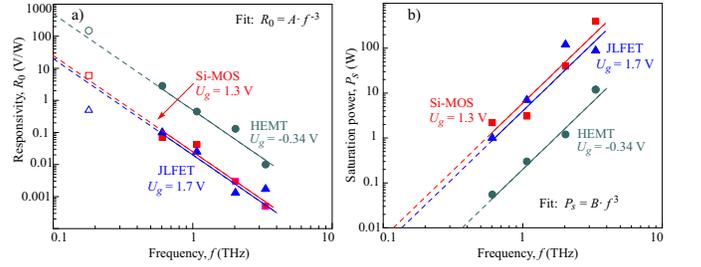}
    \caption{Frequency dependencies of: (a)  the low power
    responsivity $R_0$ being proportional to the detectors responsivity
    and (b) saturation power  $P_s$. Lines are fits after $R_0 =A f^{-3}$
    and $P_s =B f^3$. The data are obtained using TIA amplifier.
%
}
\label{fig09}
\end{figure}

Photoresponse was detected at all  frequencies ranging from 0.18
to 3.3~THz and in a wide range of frequencies and radiation power
ranging for microwatts to kilowatts. Figure~\ref{fig08}(a)  shows
power dependence of Si-MOS, JLFET and HEMT with TIA amplifiers,
i.e. in the regime of the fast response. Power dependences for the
fastest of our transistors (HEMT) with both kinds of amplifiers
are presented in Fig.~\ref{fig08}(b) for pulsed laser radiation
and Fig.~\ref{fig08}(c) for low power $cw$ laser radiation.  These
figures demonstrate that for under all conditions (different
radiation frequencies, gate voltages and type of amplifier) the
photoresponse behaves in the same way: linear response at low
powers changes to its saturation at higher powers. Furthermore, it
follows a universal phenomenological  equation:
\begin{equation}
\label{power} U = \frac{R_0 P}{(1 + P/P_s)} \: ,
\end{equation}
where $R_0 = U_0/P$ is the responsivity and $U_0$ is the signal in
the linear regime ($P\ll P_s$), and $P_s$ is the saturation power.
It is seen that both, the low power linear responsivity and
saturation power, depend strongly on the radiation frequency.
Frequency dependence of the low power responsivity $R_0$ and
saturation power $P_s$ are shown in Figs.~\ref{fig09}(a) and (b),
respectively. These plots reveal that  $R_0$ decreases with
radiation frequency as $R_0 = A f^{-3}$ while the saturation power
increases with rising frequency as $P_s = B f^3$. Studying Si-MOS
and JLFET we observed that also in these structures increase of
power leads to saturation described by Eq.~(\ref{power}) with $R_0
\propto f^{-3}$  and $P_s \propto f^3$, see Figs~\ref{fig09}.

While the observed power dependence  given by Eq.~(\ref{power}) is
in agreement with the experimental results obtained for InGaAs
HEMTs and developed broadband detection model, see~\cite{But2014},
it contradicts with the results of theoretical analysis of
Ref.~\cite{Gutin}. The latter considered nonlinear dependence of
the FET response on the power of external radiation for
non-resonant detection, when plasma oscillations are overdamped.
These regime corresponds to conditions of our experiments. The
authors demonstrated that above threshold the photoinduced voltage
behaves as:
\begin{equation}
\label{powerK1}
U = \sqrt{U_g^2 + U_a^2/2} - U_g = \frac{U_a^2}{\sqrt{U_g^2 + U_a^2/2}+U_g} \: ,
\end{equation}
and below threshold as:
\begin{equation}
\label{powerK2}
U = \frac{\eta k_BT}{e}{\rm ln} I_0\left( \frac{eU_a}{\eta k_BT}\right)\: ,
\end{equation}
where $U_a$ is the magnitude  of the radiation wave ($U_a^2\propto
P$), $k_B$ is the Boltzmann constant, $T$ is temperature, $I_0$ is
the Bessel function of the imaginary argument, and $\eta$ is the
subthreshold ideality factor. In the case of strong signal
corresponding to high power these equations predict $U \propto U_a
\propto \sqrt{P}$. Experiment of Ref.~\cite{Gutin} carried out
applying low power $cw$ molecular laser operating at frequency
1.63~THz clearly shows begin of nonlinearity and can be fitted by
Eqs.~(\ref{powerK1}) and (\ref{powerK2}). However, at used low
powers, corresponding  to a point where the nonlinearities occurr,
the data can be well fitted by Eq.~(\ref{power}), too. Analysis of
our data obtained in substantially larger range of radiation
powers demonstrates that for power levels below saturation power
the data can be well fitted for both power dependencies. At powers
above the saturation power, however, the photoresponse follows
Eq.~(\ref{power}) and can not be described by Eqs.~(\ref{powerK1})
and (\ref{powerK2}). We note that for high powers a linear
decrease of the current responsivity with increase of radiation
power has also been calculated in
Ref.~\cite{Lisauskas2018,Ikamas20182}. In these works the authors
implemented both the standard non-quasi-static FET model and the
distributed transistor model and calculated in the Advanced Design
System software environment of Keysight Technologies. As compared
to Ref.~\cite{Gutin} the authors took into account that in the
large signal approximation, the impedance starts to depend on
power as well.

Finally we discuss the frequency  dependence of the low power
responsivity and saturation power. Our data shows that signal of
all three transistors having the same antenna design and using Si
lens behaves similarly upon variation of radiation frequency
yielding $R_0 ~\propto f^3$. This result can be described in terms
of theory of Sachno et al.~\cite{Sachno2013} considering silicon
MOSFET as THz/sub-THz detectors with account of parasitic
resistances and capacitances as well as detector-antenna impedance
matching~\cite{Kopyt2015}. The work Ref.~\cite{Sachno2013} shows
that frequency dependence of real detectors depends substantially
on above parameters. It has been demonstrated that the
responsivity can decrease with the frequency increase as $R_0
\propto f^{-\nu}$ with parameter $\nu = 2$  or 4 depending on the
antenna-detector coupling.  Fitting curves in Fig.~\ref{fig09}(a)
show that for our detectors $\nu = 3$. We emphasize that, as it
has recently been demonstrated by Ikamas et al. FET detectors with
optimized antenna-transistor coupling requiring also a careful
control of beam profile, a nearly flat responsivity could be
achieved up to about 1.6 THz, see~\cite{Ikamas2018}.

As concerning the saturation power  it also scales with the third
power of radiation frequency increasing with the frequency
increase $P_s~\propto f^3$, see Fig.~\ref{fig09}(b). It is also
the smallest for HEMT, which has the largest responsivity.
Currently we have no explanation for this fact, however, we note
that in the saturation problems it is quite often that the
saturation parameter is coupled to the reciprocal value of the
radiation-matter coupling, which, in our case, is included in the
low power responsivity. At last but not least we discuss the
dynamic range of our detectors. Figure~\ref{fig09}(b) reveals that
both Si-based FETs have substantially larger (by about 20 times)
upper limit of linear detection (defined by saturation power) as
compared to HEMT. At the same time NEP of HEMT is $2\times 10^3$
times smaller than those of Si-FETs, see Tab.~\ref{tab03}.
Therefore, the best dynamic range is obtained for HEMT detectors.

\section{Conclusions}

Our study demonstrated that Si-based FET and InGaAs-based HEMT can
be used as ultrafast room temperature detectors of terahertz
radiation. Applying short pulses of spontaneously mode-locked
molecular optically pumped THz lasers we have shown that all
examined detectors equipped with low input resistance
transimpedance amplifiers had a resolution better than 4~ns. This
result reveal that applying current reading method serves
ultrafast detection of THz radiation, for HEMT, but, importantly,
for Si-FETs as well. Using of high input resistance voltage
followers decreases time resolution of Si-based MOS, so that the
time constant becomes of the order of hundreds of nanoseconds,
whereas HEMT detectors remains fast being characterized by the
time constants of the order of several nanoseconds. Investigations
of the photoresponse in a wide range of radiation powers and
frequencies demonstrated that with the power increase the linear
response changes to the signal saturation. This process is well
described by empirical equation $U\propto R_0/(1+P/P_s)$. The
range of linearity and values of saturation powers $P_s$ depends
on the radiation frequency. Our data show that the responsivity
decreases with the frequency increase as $R_0 \propto f^{-3}$,
whereas the saturation power increases  as $P_s\propto f^3$. The
latter, depending on the radiation frequency and type of
transistor changes from 0.1 up to 600~W.  To summarize, we show
that field effect transistors equipped with transimpedance
amplifiers  can be used as ultra-fast room temperature detectors
of THz radiation and that the their dynamic range extends over
many orders of magnitude of power of incoming THz radiation.
Taking into account very low NEP of HEMT we found out that while
the saturation power of this detector is the smallest it has by
orders of magnitude larger dynamic range.  Most of scientists
characterize FET-based detectors measuring the THz induced voltage
on the open drain terminal. TIA provides measurement of the THz
induced current through  the short circuited drain. The latter
configuration except high dynamic range offers also fast impulse
response and excellent signal-to-noise ratio. Therefore, we indeed
show that these FETs with current based read-out are well adopted
for operating and/or imaging with high power pulsed sources.

\begin{acknowledgements}
    We thank V. Kachorovskii and A. Lisauskas for fruitful discussions.
    Support by the CENTERA, Deutsche Forschungsgemeinschaft (DFG) and the
    Volkswagen Stiftung Program (90298) is gratefully acknowledged. This
    study was partially supported by the National Center for Research and
    Development in Poland grants LIDER/020/319/L-5/13/NCBR/2014, PBS3/B3/30/2015, PBS3/A3/18/2015.
\end{acknowledgements}

\end{document}